\documentclass[prl,aps,showpacs,twocolumn,superscriptaddress]{revtex4}
\usepackage{graphics,bm}
\usepackage{amssymb}
\usepackage{epsfig}
\usepackage{epsf}

\def\be{\begin{equation}} \def\ee{\end{equation}}
\def\bea{\begin{eqnarray}} \def\eea{\end{eqnarray}}
\def\nn{\nonumber} \def\pp{\parallel}

\begin{document}

\title{Spectroscopic Imaging Scanning Tunneling Microscopy
as a Probe of Orbital Structures and Ordering}
\author{Wei-Cheng Lee}

\affiliation{Department of Physics, University of California, 
San Diego, CA 92093}

\author{Congjun Wu}

\affiliation{Department of Physics, University of California, 
San Diego, CA 92093}

  
\begin{abstract}
Unlike charge and spin, the orbital degree of freedom of electrons 
in transition metal oxides is difficult to detect.
We present the theoretical study of a new detection method in metallic
orbitally active systems by analyzing the quasiparticle scattering 
interference (QPI) pattern of the spectroscopic imaging scanning 
tunneling spectroscopy, which is sensitive to orbital structures
and orbital ordering.
The QPIs for the $d_{xz}$ and $d_{yz}$-orbital bands in the 
$t_{2g}$-orbital systems show a characteristic stripe-like feature 
as a consequence of their quasi-one-dimensional nature, which is 
robust against orbital hybridization.
With the occurrence of orbital ordering proposed in Sr$_3$Ru$_2$O$_7$
and iron-pnictides, the stripe-like QPI patterns exhibit nematic 
distortion breaking the $C_4$-symmetry.
\end{abstract}

\maketitle

Orbital, a degree of freedom independent of charge and spin, plays an 
important role in various phenomena of transition metal oxides ($d$-orbital) 
and heavy-fermion compounds ($f$-orbital), including metal-insulator
transitions, unconventional superconductivity, colossal 
magnetoresistance \cite{imada1998,tokura2000, khaliullin2005,khomskii2005}.
Orbital ordering and excitations have been observed in many 
Mott-insulating transition metal oxides such as La$_{1-x}$Sr$_x$MnO$_3$,
La$_4$Ru$_2$O$_{10}$, LaTiO$_3$, YTiO$_3$,  
KCuF$_3$, etc. \cite{murakami1998,ulrich2008,ichikawa2000,khalifah2002}
In addition, cold atom optical lattices have opened up a new opportunity 
to study orbital physics with both bosons and fermions, which has
recently attracted considerable experimental and theoretical research 
attentions\cite{muller2007,liu2006,wu2006,wu2008,wu2008a,wu2009,lee2009a}.

Many metallic transition metal oxides, such as strontium ruthenates
and iron pnictides, are orbitally active. 
Their Fermi surfaces are composed of different components
of the $t_{2g}$-orbitals, {\it i.e.}, $d_{xy}$, $d_{xz}$ and $d_{yz}$.
Different from the $d_{xy}$-band which is quasi-two-dimensional (2D),
the $d_{xz}$ and $d_{yz}$-bands are quasi-one-dimensional with
strong in-plane anisotropy.
Their Fermi surfaces are strongly nested, resulting in 
strong incommensurate spin fluctuations in strontium
ruthenates and iron pnictides \cite{mackenzie2003,mazin2009,zhai2009}.
Furthermore, the quasi-1D bands also play an important role in the electronic 
nematic ordering observed in the bilayer Sr$_3$Ru$_2$O$_7$ \cite{grigera2001,
grigera2004,borzi2007} between two 
consecutive metamagnetic transitions in the external magnetic field,
which contributes another intriguing example of spin-orbital interplay
\cite{kee2005,puetter2007,yamase2007,tamai2008,mercure2009}.
The nematic ordering has been interpreted as orbital ordering between 
$d_{xz}$ and $d_{yz}$-orbitals by us \cite{lee2009} and also 
independently by Raghu {\it et al.} \cite{raghu2009}.

In contrast to charge and spin whose detection methods have been 
maturely developed, the orbital degree of freedom is very difficult 
to measure especially in metallic orbital systems.
In this letter, we present the theoretical study of a new method
to detect the orbital degree of freedom by employing the
technique of spectroscopic imaging scanning tunneling microscopy (SI-STM).
This technique is an important tool  
to study competing orders in strongly correlated systems
\cite{kohsaka2008,hanaguri2007,wang2003,balatsky2006,yin2009},
and has been just applied into the metallic $t_{2g}$-orbital 
systems of Sr$_3$Ru$_2$O$_7$ \cite{davis-private}.
We find that this technique provides a sensitive method to detect 
orbital degree of freedom and orbital ordering by studying
the quasi-particle interference (QPI) in the quasi-1D $d_{xz}$  
and $d_{yz}$ bands.
In contrast to the well-established QPI scenario for the single band 
system before,
the $T$-matrix acquires momentum dependent form factors
which forbid some QPI wavevectors and result in stripe features 
in the Fourier transformed STM images.
The orbital ordering exhibits in the nematic distortion of the stripe
QPI patterns.
The applications of our analysis to the nematic orbital ordering in strontium 
ruthenates and the iron pnictide superconductors will be demonstrated.

We consider the band Hamiltonian with the $d_{xz}$ and $d_{yz}$-orbital 
bands as: $H_0=\sum_{\vec{k}\sigma} H_{\vec{k}\sigma}$, and
\bea
H_{\vec{k}\sigma}&=&\epsilon_{xz,\vec{k}}d^\dagger_{xz\vec{k}\sigma} 
d_{xz,\vec{k}\sigma}+\epsilon_{yz,\vec{k}}d^\dagger_{yz,\vec{k}\sigma} 
d_{yz,\vec{k}\sigma}  \nn \\
&+& (f_{\vec{k}\sigma} d^\dagger_{xz,\vec{k}\sigma} d_{yz, \vec{k}\sigma} + h.c.),
\eea
where
$\epsilon_{xz,\vec{k}}=-2t_\pp\cos k_x-2t_\perp\cos k_y-4t'\cos k_x\cos k_y,
\epsilon_{yz,\vec{k}}=-2t_\perp\cos k_x-2t_\pp\cos k_y-4t'\cos k_x\cos k_y$.
$f_{\vec{k}\sigma}$ is the hybridization between $d_{xz}$ and $d_{yz}$ 
orbitals, which is different from materials to materials and can 
be complex function in general.
$t_{\pp}$ and $t_{\perp}$ are the nearest neighbor longitudinal and transverse 
hopping integrals for the $d_{xz}$ and $d_{yz}$-orbitals, and 
$t_{\pp} >> t_{\perp}$.
$t^\prime$ is the next-nearest neighbor intra-orbital hopping integral.
We define the basis of the pseudo-spinor as $\hat{\phi}_{\vec{k}\sigma}
=(d_{xz\vec{k}\sigma},d_{yz\vec{k}\sigma})^T$.
$H_{\vec k \sigma}$ can be diagonalized by introducing the unitary 
transformation $\hat{U}_{\vec{k}\sigma}$ such that 
$\hat{U}^\dagger_{\vec{k}\sigma} \hat{H}_{\vec{k}\sigma} \hat{U}_{\vec{k}\sigma}
={\rm diag}\{E^+_{\vec{k}\sigma},E^-_{\vec{k}\sigma}\}$.
$U$ reads in the basis of $\hat{\phi}_{\vec k \sigma}$ as
\bea
\hat{U}_{\vec{k}\sigma}=
\left(
\begin{array}{cc}
\cos\theta_{\vec{k}\sigma}&-e^{i\delta_{\vec{k}\sigma}}\sin\theta_{\vec{k}\sigma}\\
e^{-i\delta_{\vec{k}\sigma}}\sin\theta_{\vec{k}\sigma}&\cos\theta_{\vec{k}\sigma}\\
\end{array}
\right),
\eea
where
$\tan 2\theta_{\vec{k}\sigma}=\frac{2 \vert f_{\vec k\sigma}\vert} 
{\epsilon_{xz,\vec{k}}-\epsilon_{yz,\vec{k}}}$, 
$\delta_{\vec{k}\sigma}=\mbox{Arg} (f_{\vec k,\sigma})$.
The eigenvalues and the corresponding eigenvectors are: 
$E^\pm_{\vec{k}\sigma}=(\epsilon_{xz,\vec{k}}+\epsilon_{yz,\vec{k}}
\pm\sqrt{(\epsilon_{xz,\vec{k}}-\epsilon_{yz,\vec{k}})^2
+4\vert f_{\vec k\sigma}\vert^2})/2$ and
$\psi_{\vec{k}\sigma}=(\gamma_{+,\vec{k}\sigma},\gamma_{-,\vec{k}\sigma})^T
=\hat{U}^\dagger_{\vec{k}\sigma}\hat{\phi}_{\vec{k}\sigma}$, respectively.

Next we introduce the scattering Hamiltonian for the non-magnetic 
single impurity at $\vec{r}_i$. 
Assuming the isotropy of the impurity, $H_{imp}$ does not mix
$d_{xz}$ and $d_{yz}$ orbitals as
$H_{imp}=V_0\sum_{i\sigma}\big(d^\dagger_{xz,i\sigma}d_{xz,i\sigma}
+d^\dagger_{yz,i\sigma}d_{yz,i\sigma}\big)\delta_{i,\vec{r}_i}$
, where we set the impurity location $\vec{r}_i=(0,0)$ at the origin. 
In the basis of the band eigenfunction $\psi_{\vec{k}\sigma}$,
$H_{imp}$ is expressed as
\be
H_{imp}=\frac{1}{N}\sum_{\vec{k},\vec{k}^\prime,\sigma}\hat{\psi}^\dagger_{\vec{k}\sigma,a}
\hat{V}^\sigma_{\vec{k},\vec{k}^\prime;ab}\hat{\psi}_{\vec{k}^\prime\sigma,b},
\ee
where $\hat{V}^\sigma_{\vec{k},\vec{k}^\prime;ab}=V_0\left[\hat{U}^\dagger_{\vec{k}\sigma}
\hat{U}_{\vec{k}^\prime\sigma}\right]_{ab}$ is the effective scattering matrix, 
and $a,b=\pm$ are eigen-band indices.
This momentum-dependence generated by the orbital hybridization
has non-trivial consequences in the QPI spectra shown later.

The Green functions with the impurity satisfy 
\bea
\hat{G}_\sigma(\vec{k},\vec{k}^\prime)=\hat{G}_{0,\sigma}(\vec{k})
\delta_{\vec{k},\vec{k}^\prime}+\hat{G}_{0,\sigma}(\vec{k})
\hat{T}^\sigma_{\vec{k},\vec{k}^\prime} \hat{G}_{0,\sigma}(\vec{k}^\prime)
\eea
where $\hat{G}$, $\hat{G}_0$ and the $T$-matrix are $2\times 2$-matrices
in terms of band indices.
The $T$-matrix and the bare Green's functions 
$\hat{G}_{0,\sigma}(\vec{k})$ defined as:
\bea
\hat{T}^\sigma_{\vec{k},\vec{k}^\prime}=\hat{V}^\sigma_{\vec{k},\vec{k}^\prime}
+\frac{1}{N}\sum_{\vec{p}}\hat{V}^\sigma_{\vec{k},\vec{p}}\hat{G}_{0,\sigma}(\vec{p})
\hat{T}^\sigma_{\vec{p},\vec{k}^\prime},
\eea
and $[\hat{G}^{-1}_{0,\sigma}(\vec{k})]_{ab}=(\omega+i\delta-E^a_{\vec{k}\sigma})\delta_{a,b}$.

In previous theoretical analysis of QPI \cite{wang2003}, the single impurity 
$T$-matrix was simplified as momentum-independent for the single band
systems.
This simplification is no longer valid in hybridized quasi-1D bands
of $d_{xz}$ and $d_{yz}$.
In the following, we consider a square lattice containing $41\times 41$ 
sites and solve the momentum-dependent $T$-matrix numerically. 
The LDOS at energy $E$, which is proportional to the conductance ($dI/dV$) measured by the STM, 
and its Fourier transformation (FT-STM) can be calculated as
\bea
\rho(\vec{r},E)&=&-\frac{1}{N\pi}\sum_{\sigma,\vec{k},\vec{k}'}{\rm Im}\big \{ e^{-i(\vec{k}-\vec{k}')\cdot \vec{r}}\nn\\
&\times&{\rm Tr}\left[\hat{U}_{\vec{k},\sigma}\hat{G}_\sigma(\vec{k},\vec{k}',E)\hat{U}^\dagger_{\vec{k}',\sigma}\right]\big \},\nn\\
\rho(\vec{q},E)&=&\frac{1}{N}\sum_{\vec{r}} e^{-i\vec{q}\cdot\vec{r}}\rho(\vec{r},E),
\eea
Note that in all the FT-STM images presented below,
$\rho(\vec{q}=0,E)$ are removed to reveal the weaker 
QPI\cite{wang2003}, and the absolute intensities of 
$\rho(\vec{q},E)$ are plotted.

\begin{figure}
\centering\epsfig{file=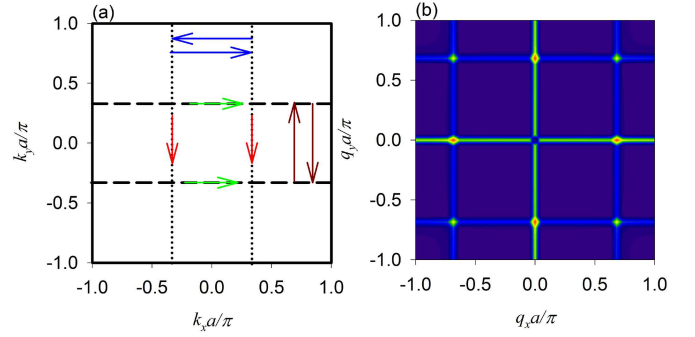}
\caption{\label{fig:toy} (a) The Fermi surfaces with 
an ideal quasi-1D bands without hybridization and (b) the corresponding 
FT-STM image. The stripe features in (b) at $q_x=0$ and $q_y=0$
result from the quasiparticle scatterings indicated by arrows within the same lines in (a), and those appearing at $q_x=\pm 2k_F$ and 
$q_y=\pm 2k_F$ come from scatterings indicated by arrows between different lines in (a), 
echoing the Friedel oscillation in exact 1D case.
}
\end{figure}

We start with a heuristic example of ideal quasi-1D case in which 
only $t_\pp$ is non-zero without hybridization.
In this case, the Fermi surface of each band is a set of two straight lines
located at $k_x=\pm k_F$ ($k_y=\pm k_F$) for $d_{xz}$ ($d_{yz}$) bands as 
shown in the Fig. \ref{fig:toy} (a).
Because the DOS is uniform along the Fermi surface, all the quasiparticle 
scatterings on the Fermi surface are equally important.
The quasiparticle scatterings occur either within the same 'Fermi lines' 
giving rise to the stripes on the $\hat{x}$ and 
$\hat{y}$ axes in the FT-STM image (Fig. \ref{fig:toy}(b)), or between 
the different 'Fermi lines' leading to the remaining weaker stripes 
in Fig. \ref{fig:toy}(b). 
These weaker stripes appearing at the lines of $q_x=\pm 2k_F$ and 
$q_y=\pm 2k_F$ are the quasi-1D analogues of Friedel oscillation in 
exact 1D systems.
Note that all the QPIs have $C_4$ symmetry because we assume that 
the $d_{xz}$ and $d_{yz}$ bands are degenerate and no spontaneous 
nematic order is present.

With turning on the hybridization, naively it may be expected that these 
stripe features should disappear since the Fermi surfaces are 2D-like.
However, we will show explicitly 
that due to the momentum-dependent $T$-matrix some quasiparticle 
scatterings on the Fermi surfaces are greatly suppressed even as the 
DOS of $\vec{k}$ points are large. 
As a result, the stripe features still survive as long as the Fermi 
surfaces remain connected.
This unique feature distinguishes the hybridized quasi-1D bands from a 
single 2D band, for example, the $d_{xy}$ band with similar Fermi 
surface topology.

\begin{figure}
\centering\epsfig{file=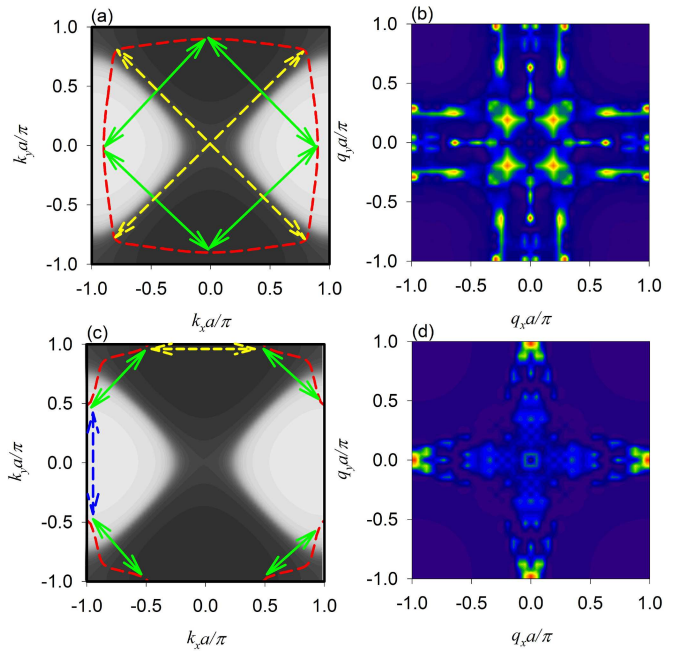}
\caption{\label{fig:sr} Fermi surfaces (dashed lines) of the two
quasi-1D bands at energies (a) just below the vH singularity ($E=1.8$)
and (c) just above the vH singularity ($E=2.0$). 
The corresponding FT-STM images are presented in (b) and (d).
The background gray scale in (a) and (c) represents the values of
$\theta_{\vec{k}}$, exhibiting from white to dark gray for
$\theta_{\vec{k}}=0\to \frac{\pi}{2} $.
The scatterings between two $\vec{k}$ points in areas with
different colors (indicated by the solid arrows) are strongly suppressed.
The stripe pattern disappears and discrete QPI wavevector points
become dominant when the Fermi surface breaks down to
discrete segments. The dashed arrows in (a) and (c) refer to the 
scatterings responsible for the strongest features in the FT-STM images.}
\end{figure}

\begin{figure}
\centering\epsfig{file=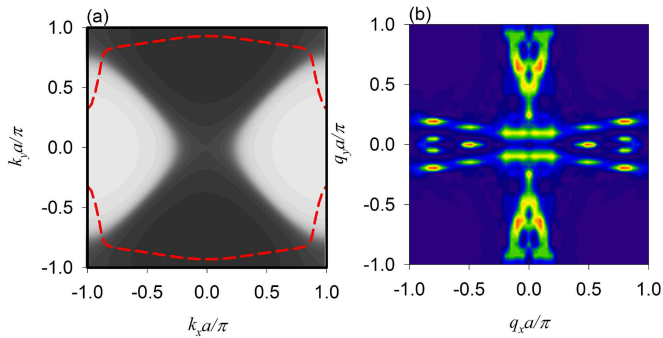}
\caption{\label{fig:nematic} (a) Fermi surface and (b) FT-STM image of the two quasi-1D band model for Sr$_3$Ru$_2$O$_7$ with nematic order right at the 
van Hove singularity ($E=1.9$).}
\end{figure}

Below we consider the on-site spin-orbit (SO) coupling
$H_{SO}=\lambda\sum_i \vec{L}_i\cdot\vec{S}_i$ to hybridize the
$d_{xz}$ and $d_{yz}$-bands \cite{eremin2002,raghu2009}.
Projecting it onto the $d_{xz}$ and $d_{yz}$-subspace, we obtain the 
hybridization function as: $f_{\vec{k}\sigma}=i\sigma\lambda/2$. 
Consequently, the effective scattering matrix $\hat{V}^\sigma_{\vec{k},
\vec{k}^\prime;ab}$ in the eigenband basis becomes:
\bea
\hat{V}^\sigma_{\vec{k},\vec{k}^\prime}= V_0
\left[
\begin{array}{cc}
\cos(\theta_{\vec{k}}-\theta_{\vec{k}^\prime})
&-i\sigma\sin(\theta_{\vec{k}}-\theta_{\vec{k}^\prime})\\
i\sigma \sin(\theta_{\vec{k}}-\theta_{\vec{k}^\prime})
&\cos(\theta_{\vec{k}}-\theta_{\vec{k}^\prime})
\end{array}
\right],
\label{vmat}
\eea
where $\tan 2\theta_{\vec{k}}=\lambda/(\epsilon_{xz,\vec{k}}-\epsilon_{yz,\vec{k}})$. 
The diagonal terms (the intra-band scattering) are  modulated by the 
form factor of $\cos(\theta_{\vec{k}}-\theta_{\vec{k}^\prime})$,
which is suppressed around $\theta_{\vec{k}}-\theta_{\vec{k}^\prime}\approx \pi/2$
is enhanced around $\theta_{\vec{k}}-\theta_{\vec{k}^\prime}\approx 0$.
For the aid to eyes, the values of the $\theta_{\vec{k}}$ are represented by
the background gray scales plotted in Figs. \ref{fig:sr}(a),(c) and \ref{fig:nematic}(a), showing white for 
$\theta_{\vec{k}}\to 0$ and dark gray for $\theta_{\vec{k}}\to \pi/2$. 
Since the larger $\hat{V}^\sigma_{\vec{k},\vec{k}^\prime}$ leads to the larger
$\hat{T}^\sigma_{\vec{k},\vec{k}^\prime}$, the QPI wavevectors connecting
two $\vec{k}$ points from different color areas have vanishing weights 
in the FT-STM images.

In the hybridized $d_{xz}$ and $d_{yz}$ bands, the DOS van Hove (vH) 
singularity occurs at $\vec X= (\pi,0)$ and $\vec X^\prime=(0,\pi)$.
Fig. \ref{fig:sr} summarizes the results for energies below and above the 
vH singularity.
The model parameters are chosen as: $(t_\pp,t_\perp,t^\prime,\lambda,V_0)
=(1.0,0.1,0.025,0.2,1.0)$ consistent with those in Ref. 
[\onlinecite{lee2009,raghu2009,eremin2002}].
In Fig. \ref{fig:sr}(b), the stripe features remain dominant in 
the FT-STM images at energy below the vH singularity as explained below. 
Although Fermi surface is a 2D closed loop shown in Fig. \ref{fig:sr} (a),
the QPI wavevectors corresponding to scattersings indicated by the 
solid arrows are prohibited due to the angular form factor discussed above.
The dominant scatterings still occur in the same way as discussed in 
Fig. \ref{fig:toy}(a), except several $\vec{q}$ vectors on the stripes 
have stronger features because of the small variations of the DOS
introduced by $t_\perp$ and $t^\prime$.
As energy crosses the vH singularity, the topology of the Fermi surface 
turns into discrete segments as shown in Fig. \ref{fig:sr}(c). 
The stripe features of the QPI wavevectors disappear and instead 
they become several discrete points whose positions depend on the model 
parameters.
As the energy is very close to the vH singularity, it has been shown
in Ref. \cite{lee2009,raghu2009} that the spontaneous nematic order
$\Delta$  appears with multi-band Hubbard interactions, which 
gives an anisotropic renormalization of dispersion of 
$\epsilon^\prime_{xz,\vec k}=\epsilon_{xz,\vec k}+\Delta$ 
and $\epsilon^\prime_{yz,\vec k}=\epsilon_{yz,\vec k}-\Delta$.
Fig. \ref{fig:nematic} plots the Fermi surface 
and the FT-STM image for the ground state with $\Delta=0.05$. 
The stripe features only extend along one particular direction and
breaks the $C_4$ symmetry down to $C_2$ symmetry, as expected for a 
nematic order.

Now we connect the above discussion to the bilayer Sr$_3$Ru$_2$O$_7$ 
system which has the additional Fermi surfaces of the  quasi-2D 
$d_{xy}$-band and the bilayer structure.
The inter-band scatterings between the quasi-2D and 1D bands are 
also suppressed due to the similar reason of different orbital
nature presented above.
The QPI pattern of the intra $d_{xy}$-band scattering should follow
the similar analysis published before \cite{wang2003,kohsaka2008}.
The quasi-1D bands of $d_{xz}$ and $d_{yz}$ have large bilayer splittings
resulting in bonding and anti-bonding versions.
Usually the impurity only lies in one layer, thus breaks the bilayer
symmetry and induces both intra and inter-band scatterings among bonding 
and anti-bonding bands.
And all of them should have the stripe pattern illustrated before.

The change of the FT-STM images as energy across the vH 
singularity can be used to distinguish the orbital configuration of 
the Fermi surface responsible for the nematic ordering observed 
in Sr$_2$Ru$_3$O$_7$, which has been proposed both in the quasi-2D 
$d_{xy}$-band \cite{kee2005,puetter2007,yamase2007}
and the quasi-1D bands of $d_{xz}$ and $d_{yz}$ \cite{lee2009,raghu2009}.
Both proposals have similiar Fermi surface topology, but the QPIs will be very different.
The stripe features are direct consequences of the quasi-1D bands 
which have comparable DOS on the Fermi surfaces.
For the 2D $d_{xy}$-bands, the QPIs are dominated by several 
discrete $\vec{q}$ vectors connecting $\vec{k}$ points with largest 
DOS as has been demonstrated nicely in the high-$T_c$ cuprate 
Bi$_2$Sr$_2$CaCu$_2$O$_{8+\delta}$\cite{kohsaka2008}.
Accordingly, we predict that if it is the 2D $d_{xy}$-band responsible 
for the nematic order, the FT-STM will show similar QPIs containing 
several discrete $\vec{q}$ vectors as the magnetic field is tuned 
through the critical point for the nematic order, while a significant 
change in the topologies of QPIs from Fig. \ref{fig:sr}(b)$\to$ Fig. \ref{fig:nematic}(b)$\to$ Fig. \ref{fig:sr}(d)
will be seen if the hybridized $d_{xz}$ and $d_{yz}$ bands are responsible.

These results may also apply to the iron pnictide superconductors
with multiple Fermi surface sheets: $\alpha_{1,2}$ bands located 
near the $\Gamma$ point composed mostly of $d_{xz}$ and $d_{yz}$-orbitals 
and $\beta_{1,2}$ bands residing near $X$ and $X'$ points with large
fraction of $d_{xy}$ orbital \cite{kuroki2008,graser2009}. 
Given that the tunneling rate along the $\hat{z}$ direction is strongly
suppressed with the increase of magnitude of in-plane momentum 
$\vert \vec{k}_\pp\vert$ \cite{tersoff1983}, the tunneling matrix 
elements of $\beta_{1,2}$ bands are naturally to be much smaller than 
those of $\alpha_{1,2}$ bands. 
The similar suppression of tunneling matrix elements at large in-plane
momentum has been demonstrated in the graphene systems \cite{zhang2008}.
As a result, SI-STM is expected to observe mostly the QPI scatterings 
from the $\alpha_{1,2}$ bands, and therefore the stripe features should 
be observed with a length roughly the size of the $\alpha_{1,2}$ 
pockets in the normal state of the iron pnictides.
More interestingly, it has been suggested \cite{zhao2009} based on a 
recent neutron scattering measurement performed on the undoped CaFe$_2$As$_2$
that a Heisenberg model with highly anisotropic in-plane exchange 
interactions is required to fit the spin-wave dispersion, 
indicating the possibility of nematic order \cite{singh2009}. 
Besides, the nematic order in LaOFeAs compound has also been 
theoretically predicted\cite{fang2008,xu2008,zhai2009}.
If such nematic order exists, the stripe features 
along one certain direction resembling Fig. \ref{fig:nematic}(b) 
should be observable in the FT-STM image.

In conclusion, we have performed the theoretical investigation of a new 
detection method to the orbital degree of freedom and the orbital ordering
in metallic transition metal oxides. 
For the quasi-1D $d_{xz}$ and $d_{yz}$ bands in the $t_{2g}$-orbital systems,
the Fourier transformed STM image of the QPIs exhibit the stripe pattern.
When the orbital hybridization is present, the $T$-matrix becomes 
momentum-dependent even for a single impurity problem and will 
suppress some QPI wavevectors depending on the hybridization 
angle $\theta_{\vec{k}}$. 
The consequences of the orbital ordering in Sr$_3$Ru$_2$O$_7$ 
and the iron pnictide superconductors have been pointed out
as a nematic distortion of the stripe pattern of the QPI.

We are grateful to J. C. Davis for his experiment results before 
publication and helpful discussion.
This work is supported by ARO-W911NF0810291
and Sloan Research Foundation.


\end{document}